\documentclass{article}
\usepackage{spconf,amsmath,graphicx,hyperref}
\usepackage{amssymb,amsfonts}
\usepackage{booktabs}
\usepackage{caption}
\usepackage{multirow}
\usepackage{siunitx}
\usepackage{subcaption}
\usepackage{float}

\newcommand{\PRB}{PRB}
\newcommand{\MSE}{MSE}
\newcommand{\MAE}{MAE}
\newcommand{\CRPS}{CRPS}
\newcommand{\TFT}{TFT}
\title{Goal-Oriented Probabilistic Forecasting for Dynamic PRB Allocation in 5G Networks}
\name{Oier Larumbe-Lizarraga$^{\star}$ \qquad Roberto Pereira$^{\star\dagger}$ \qquad Cristian J. Vaca-Rubio$^{\ddagger}$\thanks{This work was performed before C. J. Vaca-Rubio joined Ericsson.}}

\address{
  $^{\star}$Universidad Isabel I, Burgos, Spain \\
  $^{\dagger}$Keysight Technologies, Barcelona, Spain \\
  $^{\ddagger}$Ericsson Research, Stockholm, Sweden \\
  \small\texttt{oier.larumbe@alumnos.ui1.es}, \texttt{robertomatheus.pinheiro@ui1.es}, \texttt{cristian.vaca.rubio@ericsson.com}
}

\begin{document}
%
\maketitle

\begin{abstract}
Efficient physical resource block (PRB) allocation in 5G networks requires accurate demand forecasting. Conventional methods minimize symmetric error metrics (MAE, RMSE), ignoring the operational cost asymmetry where under-provisioning (service degradation) is far costlier than over-provisioning (wasted capacity). We propose a goal-oriented probabilistic forecasting framework that aligns model training with the operator's decision-making objectives. Specifically, we train DeepAR and temporal fusion transformer (TFT) models using the Pinball Loss function and derive the optimal allocation quantile from the operator's cost matrix. Evaluation on a real beam-level 5G traffic dataset shows that the proposed approach reduces operational cost compared to MSE-trained baselines while maintaining calibrated uncertainty estimates. The framework enables dynamic PRB allocation that explicitly balances service reliability against resource efficiency.
\end{abstract}
\begin{figure*}[t]
\centering
\includegraphics[width=\textwidth]{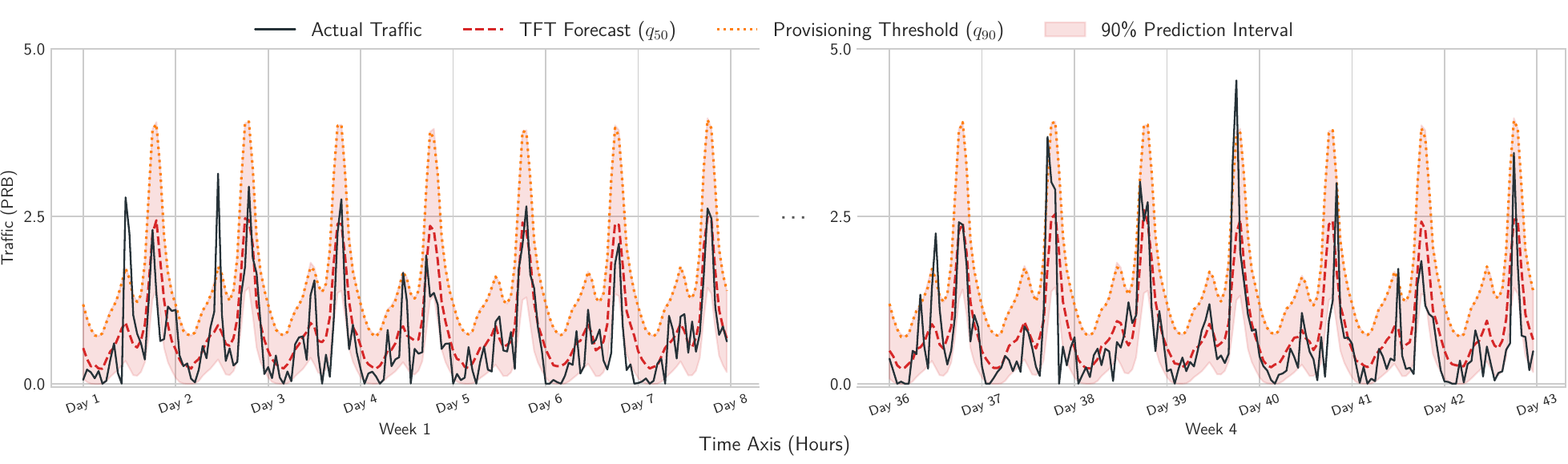}
\caption{TFT probabilistic forecasts ($q_{90}$) for the high-traffgic stratum during early and late validation periods.
}
\label{fig:tft_high_stratum}
\end{figure*}
\section{Introduction}
\label{sec:intro}
In 5G and beyond-5G networks, physical resource block (PRB) allocation at the beam level directly governs the capacity available to users \cite{wang2024survey}. Because traffic demand at this granularity is highly dynamic, driven by user activity, mobility, and spatial load variations, reliable forecasting is essential for proactive resource management and early detection of critical load conditions.


Traditional time series forecasting approaches, from classical statistical models such as ARIMA ~\cite{arima} to deep learning solutions such as LSTMs ~\cite{lstm}, are typically trained by minimizing symmetric prediction losses such as mean squared error (MSE) or mean absolute error (MAE). These objectives implicitly treat positive and negative forecast errors as equally undesirable. 
However, in practical telecommunication operations, 
these two errors may lead to different consequences. 
Underestimating the PRB demand may lead to insufficient capacity (hereafter referred to as under-provisioning), degrading the quality of service and potentially violating service level agreements~\cite{zervou2025spatio}. Conversely, overestimating PRB demand (hereafter referred to as over-provisioning) will lead to unused capacity and increase energy consumption~\cite{ lopez}. Because symmetric loss functions do not encode this asymmetry, models optimized for MAE or MSE are not necessarily the optimal choice for operational tasks.


Probabilistic forecasting models such as DeepAR~\cite{DeepAR} and the Temporal Fusion Transformer (TFT)~\cite{TFT}
address part of the problem by 
estimating predictive distributions rather than single point estimates, which in turn may enable quantile-based decision rules for resource provisioning ~\cite{kasuluru2026ai}.
%
%
However, producing a calibrated predictive distribution alone does not guarantee that the forecasting model is optimized for the downstream network objective \cite{elmachtoub2022smart}.
Decision-focused learning (DFL)~\cite{shah2022decisionfocused} tries to address this gap by embedding the downstream decision problem directly into the training loop.
Classical DFL formulations, however, often require differentiating through, or repeatedly solving/approximating, the downstream optimization problem.
%
Prior work on goal-oriented forecasting for network resources  ~\cite{CristianRoberto, mootoo2025emforecaster, salem2025goaloriented} has begun to explore this direction, 
but has not established an explicit 
link between the operator's asymmetric provisioning costs and the training objective.


In this work, we exploit the simple structure of PRB provisioning to establish exactly that link: modeling under- and over-provisioning directly through distinct penalty coefficients, we show that the cost-minimizing allocation is a conditional quantile whose level follows analytically from these penalties
This maps the operational cost directly onto pinball loss~\cite{pinball}, so the downstream objective enters training without an explicit optimization layer. Our contributions are threefold: (i) we derive a closed-form optimal decision quantile from asymmetric under-/over-provisioning penalties, providing a lightweight alternative to classical DFL that avoids differentiating through the downstream allocation problem; (ii) we instantiate this framework on
   DeepAR and TFT, benchmarked against deterministic and Gaussian-likelihood baselines on a real-world beam-level 5G traffic dataset over short- and long-term horizons; and (iii) we show that pinball-trained TFT cuts operational cost by up to 43.6\% relative to MSE-trained baselines in the short-term
  case.

\section{System Model and Problem Formulation}
\label{sec:system}

The \PRB{} utilization of a given beam at time step $t$ is modeled as a discrete-time stochastic process, denoted as $\{y_t\}_{t \in \mathbb{Z}^+}$, where $y_t \in [0, 1]$ represents the normalized capacity required to satisfy user traffic demand. The objective of probabilistic forecasting is to estimate the conditional probability density function of future \PRB{} demand over a multi-step forecasting horizon $H$, given a trajectory of $t$ past observations. Formally, we aim to model:
\begin{equation}
    p(y_{t+1:t+H} \mid y_{1:t}) = \prod_{h=1}^{H} p(y_{t+h} \mid y_{1:t+h-1}),
\end{equation}
where $h \in \{1, \dots, H\}$ denotes the forecasting step. Rather than outputting a single point estimate $\hat{y}_{t+h}$, the probabilistic model predicts the complete predictive distribution, enabling the extraction of target quantiles $q_{\tau}(y_{t+h} \mid y_{1:t})$ associated with a specific coverage probability $\tau \in (0, 1)$.
\subsection{Asymmetric Operational Cost}
\label{ssec:cost}
Operational costs in \PRB{} allocation are inherently asymmetric. Let $y_t$ denote the actual \PRB{} demand and $A_t$ the allocated \PRB{} capacity at time step $t$, the under- and over-provisioning rates are given by $U = \max(0, y_{t} - A_{t})$ and $O = \max(0, A_{t} - y_{t})$, respectively. Under-provisioning directly degrades 
QoS 
and incurs severe service level agreement penalties, whereas over-provisioning results in energetic waste. Consequently, the total operational cost $C$ is modeled as a linear combination of both components:
\begin{equation}
\label{eq:cost}
    C = c_u \cdot U + c_o \cdot O,
\end{equation}
where $c_u$ and $c_o$ represent the under- and over-provisioning penalty coefficients, respectively, with $c_u \gg c_o$. Conventional forecasting models are typically trained by minimizing \MSE{}, a symmetric metric that fails to reflect this operational asymmetry \eqref{eq:cost}. To address this limitation, we adopt the pinball loss~\cite{pinball} function:
\begin{equation}
\label{eq:PB}
	L_\tau(y_t, \hat{y}_t) = \tau \cdot \max(0, y_t - \hat{y}_t) + (1 - \tau) \cdot \max(0, \hat{y}_t - y_t).
\end{equation}
Unlike MSE, the pinball loss applies asymmetric slopes to prediction residuals, weighting under-predictions ($y_t > \hat{y}_t$) by $\tau$ and over-predictions ($y_t \le \hat{y}_t$) by $(1-\tau)$. 
Consequently, minimizing $L_\tau$ yields the conditional quantile $q_\tau(y_t)$ rather than a central mean expectation. Comparing \eqref{eq:cost} and \eqref{eq:PB} reveals a direct structural
correspondence: when setting the allocated capacity $A_t = \hat{y}_t$, the operational cost $C$ becomes proportional to the pinball loss $L_\tau$ up to the positive constant $(c_u + c_o)$, with penalty parameters $c_u$ and $c_o$ mapping to the  weights $\tau$ and $(1-\tau)$, respectively. Since minimizing a positive rescaling of a function is equivalent to minimizing the function itself, parameterizing $\tau$ via $c_u$ and $c_o$ directly aligns model training with the operator's 
objective.
\subsection{Optimal Quantile Derivation}
\label{ssec:quantile}
  Setting $A_t = \hat{y}_t$ in~\eqref{eq:cost} and comparing term by term
  with~\eqref{eq:PB}, we seek a quantile level $\tau^*$ and a positive
  constant $k$ such that $C = k \cdot L_{\tau^*}$. Expanding both sides:
  \begin{equation}
    k\,\tau^* = c_u, \qquad k\,(1 - \tau^*) = c_o.
    \label{eq:system}
  \end{equation}
  Dividing the first expression by the second eliminates $k$
  %
  %
  and solving for $\tau^*$, yields the closed-form optimal decision:
  \begin{equation}
    \tau^* = \frac{c_u}{c_u + c_o},
    \label{eq:tau_star}
  \end{equation}
  with the proportionality constant $k = c_u + c_o$.
  This result guarantees that training a probabilistic model with the pinball
  loss at level $\tau^*$ directly minimizes the operator's expected
  operational cost~\eqref{eq:cost}, without requiring an explicit
  optimization layer or differentiation through the downstream allocation
  problem.


\section{Proposed Methodology}
\label{sec:method}
\begin{table*}[t]
\centering
\caption{Comprehensive Evaluation: Statistical, Calibration, Over/Under-provisioning, and Operational Performance Across Forecasting Horizons.}
\label{tab:master_results_3digits}
\footnotesize
\setlength{\tabcolsep}{1.9pt}
\begin{tabular}{ll cccccccc c cccccccc}
\toprule
 & & \multicolumn{8}{c}{\textbf{Short-Term Horizon ($V_1$)}} & & \multicolumn{8}{c}{\textbf{Long-Term Horizon ($V_2$)}} \\
\cmidrule{3-10} \cmidrule{12-19}
\textbf{Model} & \textbf{Loss} & \textbf{MAE} & \textbf{CRPS} & \textbf{Cov.} & \textbf{Over.} & \textbf{Under.} & \textbf{Cost} & \textbf{Sleep} & \textbf{Elast.} & & \textbf{MAE} & \textbf{CRPS} & \textbf{Cov.} & \textbf{Over.} & \textbf{Under.} & \textbf{Cost} & \textbf{Sleep} & \textbf{Elast.} \\
\midrule
\multicolumn{19}{l}{\textit{Deterministic Baselines (Point Prediction $\mu$)}} \\
\quad SARIMA & MSE & 0.254 & -- & -- & 0.139 & 0.114 & 0.117 & \textbf{28.5\%} & 63.6\% & & 0.358 & -- & -- & \textbf{0.210} & 0.149 & 0.155 & \textbf{27.5\%} & \textbf{61.4\%} \\
\quad LSTM Det. & MSE & 0.267 & -- & -- & \textbf{0.133} & 0.133 & 0.133 & 27.8\% & \textbf{64.5\%} & & 0.616 & -- & -- & 0.524 & 0.092 & 0.135 & 1.0\% & 37.4\% \\
\midrule
\multicolumn{19}{l}{\textit{Probabilistic Models (Quantile Policy $q_{90}$)}} \\
\quad DeepAR & Gaussian & 0.262 & 0.191 & 88.8\% & 0.443 & 0.044 & 0.084 & 1.1\% & 46.7\% & & \textbf{0.340} & \textbf{0.247} & 90.6\% & 0.710 & 0.048 & \textbf{0.114} & 2.7\% & 36.0\% \\
\quad DeepAR & Pinball & 0.266 & 0.192 & 90.3\% & 0.556 & \textbf{0.029} & 0.082 & 0.0\% & 41.3\% & & 0.461 & 0.331 & 82.0\% & 0.971 & \textbf{0.027} & 0.121 & 0.0\% & 27.6\% \\
\quad LSTM Prob. & Gaussian & 0.277 & 0.200 & \textbf{93.1\%} & 0.500 & 0.034 & 0.081 & 0.1\% & 40.8\% & & 0.420 & 0.311 & \textbf{91.0\%} & 0.649 & 0.082 & 0.139 & 0.0\% & 31.3\% \\
\quad TFT & Pinball & \textbf{0.241} & \textbf{0.177} & 82.0\% & 0.440 & 0.034 & \textbf{0.075} & 0.0\% & 44.2\% & & 0.371 & 0.268 & 71.8\% & 0.691 & 0.052 & 0.116 & 0.0\% & 31.5\% \\
\bottomrule
\end{tabular}
\end{table*}
\subsection{Pinball Loss for Goal-Oriented Training}
\label{ssec:pinball}
We train DeepAR and \TFT{} with the pinball loss, setting $\tau = \tau^*$ so that the resulting probabilistic forecasts explicitly optimize the operator's asymmetric cost matrix, unlike models trained under standard \MSE{} loss.
For a given observation $y_t$, the quality of the predicted distribution $F$ is evaluated using the continuous ranked probability score (CRPS), which relates directly to the pinball loss  over all forecast quantiles:
\begin{equation}
    \text{CRPS}(F, y_t) = 2 \int_{0}^{1} L_{\tau}(y_t, q_{\tau}) \, d\tau.
\end{equation}
In the well-specified limit, minimizing the expected pinball loss guarantees $\mathbb{P}(y_t \le q_\tau) = \tau$; in practice, finite-sample and model-misspecification effects may lead to deviations from nominal coverage. 
\subsection{Probabilistic Forecasting Models}
\label{ssec:models}
  We benchmark two probabilistic architectures, DeepAR and TFT, against
  deterministic baselines (SARIMA and a point-prediction LSTM trained with MSE).

  DeepAR~\cite{DeepAR} is an autoregressive recurrent
  network that generates forecasts sequentially: at each horizon step~$h$, the
  decoder conditions on all previous predictions $\hat{y}_{t+1:t+h-1}$.
  In its standard formulation, the output layer parametrizes the mean~$\mu$
  and standard deviation~$\sigma$ of a Gaussian likelihood, from which
  quantiles are derived analytically. In the pinball variant, we replace the
  Gaussian head with a direct quantile output trained under the pinball loss
  at level~$\tau^*$, so that the network directly minimizes the operator's
  asymmetric cost.

  TFT~\cite{TFT} is a transformer-based architecture that
  differs from DeepAR in two key aspects. First, it produces all $H$~horizon
  steps in parallel via a multi-horizon forecasting head, rather than
  autoregressively. Second, it incorporates a variable selection network and
  interpretable self-attention over past time steps. The output layer emits
  multiple quantiles simultaneously, trained under a multi-quantile pinball
  loss that includes~$\tau^*$ among the target levels.
\section{Experimental Results}
  \label{sec:results}

  \subsection{Dataset and Experimental Setup}
  \label{ssec:dataset}

  The dataset comprises anonymized, curated real-world 5G traffic data
  measured at the beam level~\cite{salem2025goaloriented}. Each base station has three
  cells of 32~beams each, for a total of 2,880~beams with hourly temporal
  granularity. Beams are grouped into three traffic strata (high, medium,
  and low) based on their average PRB utilization.

  Model training uses a stratified sample of 150~beams (50 per stratum)
  over Weeks~1--5. Evaluation is performed on a disjoint set of 450~beams
  (150 per stratum) under an expanding-window scheme across two validation
  periods: a short-term horizon~($V_1$, Week~6) immediately following
  training, and a long-term horizon~($V_2$, Week~11) separated by a
  four-week gap to assess robustness to temporal drift
  (see Fig.~\ref{fig:tft_high_stratum}). The forecasting horizon is
  $H = 168$~hours (one week). All probabilistic models use a quantile policy at
  level $\tau^* = c_u/(c_u + c_o)$ with $c_u/c_o = 9$ (i.e., $\tau^* = 0.9$), reflecting an operator profile where under-provisioning is costlier than over-provisioning.

  Performance is evaluated along three axes:
  \emph{(i)~Statistical accuracy}: \MAE{} for point forecasts
  (deterministic baselines and median predictions~$q_{50}$ of
  probabilistic models) and \CRPS{}~\cite{crps} for full predictive
  distributions.
  \emph{(ii)~Calibration}: empirical coverage~\cite{coverage} of the 90\%
  prediction interval.
  \emph{(iii)~Operational impact}: under- and over-provisioning
  rates~($U$,~$O$) and total operational cost, together with two
  energy-efficiency indicators: elastic capacity savings (PRB allocation 
reduction relative to a static maximum normalized baseline $M = 1.0$) and cell-sleep savings (fraction
of time steps where allocated capacity $A_t$ falls below a 10\% threshold, i.e., $A_t \le 0.10$, enabling cell deactivation).
\subsection{Forecasting Performance}
  \label{ssec:performance}

  Table~\ref{tab:master_results_3digits} reports forecasting performance
  across both validation horizons. In~$V_1$, \TFT{} Fig.1., achieves the lowest
  \CRPS{} among all probabilistic models, followed by Gaussian
  DeepAR. We attribute this advantage to the multi-horizon
  forecasting head of \TFT{}, which predicts all $H$~steps in parallel
  and avoids the autoregressive error accumulation that affects DeepAR.
  Comparing median predictions~($q_{50}$) against SARIMA point forecasts
  confirms this: \TFT{} reduces \MAE{} by 5.1\%, a difference that is statistically significant.

  In~$V_2$, however, Gaussian DeepAR overtakes \TFT{} in both \CRPS{}
   and \MAE{}. This reversal
  reflects the higher sensitivity of pinball-trained models to
  distribution shift: because the pinball loss tightly fits the quantile
  structure observed during training, any change in the traffic
  distribution between $V_1$ and $V_2$ displaces the learned quantiles.
  Gaussian-likelihood models, by contrast, parametrize $\mu$ and
  $\sigma$ of a smooth distribution and degrade more gracefully.

  Regarding calibration, Gaussian models maintain empirical coverage
  close to the 90\% nominal level across both horizons. \TFT{}, however, exhibits
  under-coverage because it trains
  on discrete quantile targets ($q_{10}$, $q_{50}$, $q_{90}$) rather
  than a continuous parametric distribution: if the learned $q_{90}$
  underestimates the true quantile, no distributional constraint
  self-corrects. This trade-off may be acceptable when the provisioning
  decision relies on a single quantile rather than the full interval.

  \subsection{Operational Impact Analysis}
  \label{ssec:impact}

  Comparing both DeepAR variants isolates the effect of the training
  loss. In~$V_1$, pinball-trained DeepAR reduces under-provisioning relative to its Gaussian counterpart,
  yielding lower operational cost. This confirms the
  core mechanism of our framework: aligning the loss with the asymmetric
  cost structure directly reduces under-provisioning, which dominates
  the cost when $c_u \gg c_o$.

  \TFT{} achieves the lowest operational cost in~$V_1$, a
  43.6\% reduction relative to the strongest deterministic baseline
  (LSTM Det.). This gain combines superior forecasting accuracy
  (lowest \CRPS{}) with an asymmetry-aware loss that shifts the
  $q_{90}$ output upward just enough to absorb demand peaks without
  excessive over-provisioning. In~$V_2$, pinball-trained models degrade more: DeepAR Pinball reaches
  an over-provisioning rate nearly double that of Gaussian
  DeepAR. Because the learned quantiles no longer match the
  shifted traffic distribution, the model over-allocates as a defensive
  response. Gaussian DeepAR achieves the lowest cost in~$V_2$,
  suggesting that symmetric likelihoods are a safer fallback when
  periodic retraining is not feasible.

  From an energy-efficiency perspective, deterministic baselines achieve
  the highest cell-sleep savings because
  their mean predictions frequently fall below the deactivation
  threshold. Pinball-trained models suppress this by design: the
  asymmetric penalty drives allocated capacity above the threshold in
  nearly all time steps, yielding 0\% cell-sleep savings. This is not a
  deficiency but an explicit consequence of prioritizing service
  reliability ($c_u/c_o = 9$). The elastic capacity metric captures the
  complementary view: \TFT{} achieves 44.2\% elastic savings in~$V_1$,
  meaning its $q_{90}$ allocation tracks demand tightly relative to
  static maximum allocation while maintaining near-zero
  under-provisioning.


  \begin{figure}[htbp]
  \centering
  \includegraphics[width=0.98\linewidth]{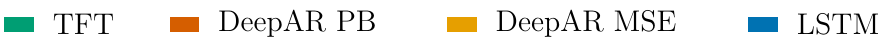}\\[0.5ex]
  \begin{subfigure}{0.5\linewidth}
      \centering
      \includegraphics[width=0.98\linewidth]{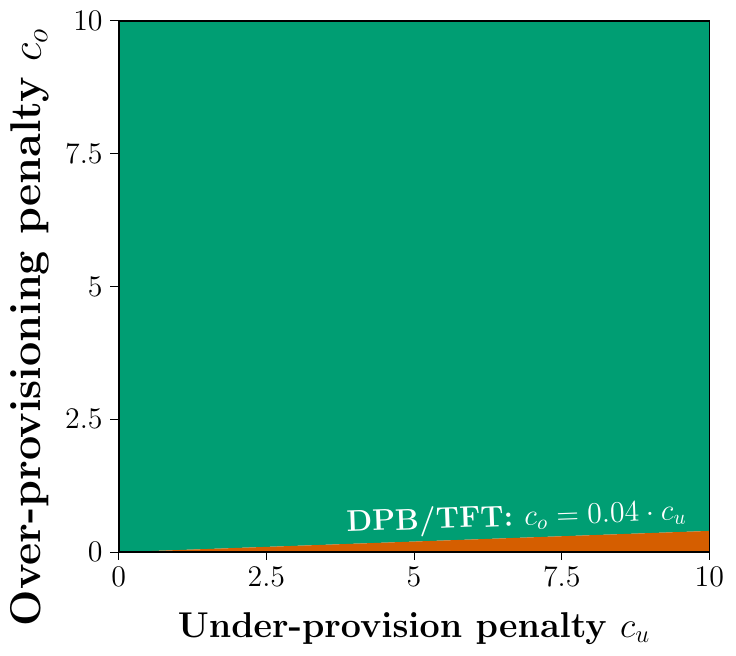}
      \caption{Short-term horizon ($V_1$)}
      \label{fig:cost_v1}
  \end{subfigure}%
  \begin{subfigure}{0.5\linewidth}
      \centering
      \includegraphics[width=0.98\linewidth]{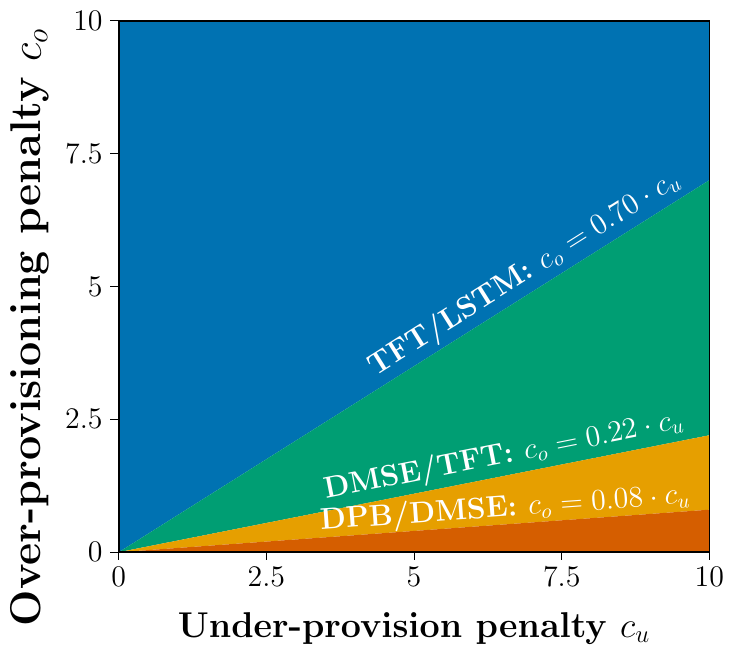}
      \caption{Long-term horizon ($V_2$)}
      \label{fig:cost_v2}
  \end{subfigure}
  \caption{Optimal model selection as a function of the under-provisioning
  penalty $c_u$ (x-axis) and over-provisioning penalty $c_o$ (y-axis).
  Colored regions indicate the model achieving the lowest operational
  cost $C$ for that penalty combination.
  (a)~Short-  ($V_1$) and (b)~Long-term horizon ($V_2$).}
  \label{fig:cost_sensitivity_both}
  \end{figure}

Finally, Fig.~\ref{fig:cost_v1} shows that \TFT{} dominates across
  nearly the entire $(c_u, c_o)$ space in~$V_1$, ceding to DeepAR
  Pinball only when $c_o < 0.04 \cdot c_u$, i.e.\ when the operator
  tolerates substantial over-provisioning.  In~$V_2$
  (Fig.~\ref{fig:cost_v2}), distribution shift reshapes the decision
  map: LSTM becomes optimal for $c_o > 0.79\,c_u$, Gaussian DeepAR
  takes over in the mid-range ($0.08\,c_u < c_o < 0.22\,c_u$), and
  DeepAR Pinball dominates the low-$c_o$ corner.  For the operational
  point considered in this work ($c_u/c_o = 9$), \TFT{} with periodic
  retraining remains the recommended choice.

\vspace{-0.5cm}
\section{Conclusion}
\label{sec:conclusion}


This paper has proposed a goal-oriented probabilistic forecasting framework for dynamic \PRB{} allocation in 5G networks. By analytically deriving the optimal decision quantile $\tau^*$ and aligning the pinball loss with the operator's asymmetric cost matrix, the proposed approach reduces operational costs by up to 43.6\% over standard \MSE{}-trained baselines. Experimental evaluation on real-world beam-level traffic data confirms that \TFT{} optimized via pinball loss achieves the best trade-off between service reliability and capacity efficiency in short-term horizons. Future work will extend this framework to multi-cell environments and online quantile adaptation.

\section*{Compliance with Ethical Standards}
  This is a numerical simulation study for which no ethical approval was required.

 \section*{Conflicts of Interest}
 No external funding was received. The authors have no other relevant financial or nonfinancial interests to disclose.


\bibliographystyle{IEEEbib}
\bibliography{strings,refs}

@article{TFT,
  title={Temporal fusion transformers for interpretable multi-horizon time series forecasting},
  author={Lim, Bryan and Ar{\i}k, Sercan {\"O} and Loeff, Nicolas and Pfister, Tomas},
  journal={International journal of forecasting},
  volume={37},
  number={4},
  pages={1748--1764},
  year={2021},
  publisher={Elsevier}
}

@article{DeepAR,
  title={DeepAR: Probabilistic forecasting with autoregressive recurrent networks},
  author={Salinas, David and Flunkert, Valentin and Gasthaus, Jan and Januschowski, Tim},
  journal={International journal of forecasting},
  volume={36},
  number={3},
  pages={1181--1191},
  year={2020},
  publisher={Elsevier}
}

@article{coverage,
	title={Evaluating interval forecasts},
	author={Christoffersen, Peter F},
	journal={International Economic Review},
	pages={841--862},
	year={1998},
	publisher={JSTOR}
}

@book{arima,
  title={Time series analysis: forecasting and control},
  author={Box, George EP and Jenkins, Gwilym M and Reinsel, Gregory C and Ljung, Greta M},
  year={2015},
  publisher={John Wiley \& Sons}
}

@article{lstm,
  title={Long short-term memory},
  author={Hochreiter, Sepp and Schmidhuber, J{\"u}rgen},
  journal={Neural computation},
  volume={9},
  number={8},
  pages={1735--1780},
  year={1997},
  publisher={MIT press}
}

@article{pinball,
  title={Regression quantiles},
  author={Koenker, Roger and Bassett Jr, Gilbert},
  journal={Econometrica: journal of the Econometric Society},
  pages={33--50},
  year={1978},
  publisher={JSTOR}
}

@article{crps,
	title={Strictly proper scoring rules, prediction, and estimation},
	author={Gneiting, Tilmann and Raftery, Adrian E},
	journal={Journal of the American Statistical Association},
	volume={102},
	number={477},
	pages={359--378},
	year={2007},
	publisher={Taylor \& Francis}
}

@inproceedings{salem2025goaloriented,
  title={Goal-oriented time-series forecasting: foundation framework design},
  author={Fechete, Luca-Andrei and Sana, Mohamed and Ayed, Fadhel and Piovesan, Nicola and Li, Wenjie and De Domenico, Antonio and Salem, Tareq Si},
  booktitle={Proceedings of the AAAI Conference on Artificial Intelligence},
  volume={40},
  number={25},
  pages={21065--21073},
  year={2026}
}

@ARTICLE{CristianRoberto,
	author={Vaca-Rubio, Cristian J. and Kasuluru, Vaishnavi and Zeydan, Engin and Blanco, Luis and Pereira, Roberto and Caus, Marius and Dev, Kapal},
	journal={IEEE Communications Standards Magazine}, 
	title={Probabilistic Forecasting for Network Resource Analysis in Integrated Terrestrial and Non-Terrestrial Networks}, 
	year={2025},
	volume={9},
	number={2},
	pages={72-79},
	doi={10.1109/MCOMSTD.2025.3568859}}

@ARTICLE{lopez,
	author={López-Pérez, David and De Domenico, Antonio and Piovesan, Nicola and Xinli, Geng and Bao, Harvey and Qitao, Song and Debbah, Mérouane},
	journal={IEEE Communications Surveys \& Tutorials}, 
	title={A Survey on 5G Radio Access Network Energy Efficiency: Massive MIMO, Lean Carrier Design, Sleep Modes, and Machine Learning}, 
	year={2022},
	volume={24},
	number={1},
	pages={653-697},
	doi={10.1109/COMST.2022.3142532}}

@article{elmachtoub2022smart,
  title={{Smart “Predict, then Optimize”}},
  author={Elmachtoub, Adam N and Grigas, Paul},
  journal={Management science},
  volume={68},
  number={1},
  pages={9--26},
  year={2022},
  publisher={INFORMS}
}

@article{shah2022decisionfocused,
  title={Decision-focused learning without decision-making: Learning locally optimized decision losses},
  author={Shah, Sanket and Wang, Kai and Wilder, Bryan and Perrault, Andrew and Tambe, Milind},
  journal={Advances in Neural Information Processing Systems},
  volume={35},
  pages={1320--1332},
  year={2022}
}

@article{wang2024survey,
  author  = {Wang, X. and Wang, Z. and Yang, K. and Song, Z. and Bian, C. and Feng, J. and Deng, C.},
  title   = {A survey on deep learning for cellular traffic prediction},
  journal = {Intelligent Computing},
  volume  = {3},
  pages   = {0054},
  year    = {2024}
}

@article{mootoo2025emforecaster,
  author  = {Mootoo, X. and Tabassum, H. and Chiaraviglio, L.},
  title   = {{EMForecaster}: A Deep Learning Framework for Time Series Forecasting in Wireless Networks With Distribution-Free Uncertainty Quantification},
  journal = {IEEE Transactions on Network Science and Engineering},
  volume  = {13},
  pages   = {1207--1225},
  year    = {2025}
}

@article{kasuluru2026ai,
  title={Ai-empowered multivariate probabilistic forecasting: A key enabler for sustainability in open ran},
  author={Kasuluru, Vaishnavi and Blanco, Luis and Vaca-Rubio, Cristian J and Zeydan, Engin and Bel, Albert},
  journal={IEEE Transactions on Network and Service Management},
  year={2026},
  publisher={IEEE}
}

@inproceedings{zervou2025spatio,
  title={A Spatio-Temporal Graph Neural Network-Based Framework for SLA Compliance in 6G Networks},
  author={Zervou, Poulcheria and Keramidi, Irene and Ramantas, Kostas and Verikoukis, Christos},
  booktitle={GLOBECOM 2025-2025 IEEE Global Communications Conference},
  pages={817--822},
  year={2025},
  organization={IEEE}
}
\nocite{*}
\end{document}